\input{aipcheck}

\documentclass[
  ,draft            
  ]
  {aipproc}

\layoutstyle{8x11single}
\usepackage{amsmath,amssymb}
\def\alt{\lesssim}

\begin{document}

\title{A New Approach based on Langevin type Equation 
for Driven Granular Gas under Gravity
}


\author{J. Wakou}{
  address={Miyakonojo National College of Technology, Miyakonojo-shi, Miyazaki, 885-8567}
}

\author{M. Isobe}{
  address={Graduate School of Engineering, Nagoya Institute of Technology, Nagoya 466-8555}
}

\begin{abstract}
We propose a novel approach based on a Langevin equation for fluctuating motion of the center of mass of granular media fluidized by energy injection from a bottom plate. In this framework, the analytical solution of the Langevin equation is used to derive analytic expressions for several macroscopic quantities and the power spectrum for the center of mass. In order to test our theory, we performed event-driven molecular dynamics simulations for one- and two-dimensional systems. Energy is injected from a vibrating bottom plate in the one-dimensional case and from a thermal wall at the bottom in the two-dimensional case. We found that the theoretical predictions are in good agreement with the results of those simulations under the assumption that the fluctuation-dissipation relation holds in the case of nearly elastic collisions between particles. However, as the inelasticity of the interparticle collisions increases, the power spectrum for the center of mass obtained by the simulations gradually deviates from the prediction of theoretical curve. Connection between this deviation and violation of the fluctuation-dissipation relation is discussed.
\end{abstract}

\maketitle


\section{Introduction}
Granular materials fluidized by external driving have been widely studied in recent years in the field of nonequilibrium statistical physics. 
In order to fluidize granular matter, we have to supply sufficiently large energy continuously, because the kinetic energy of grains is dissipated due to inelastic collisions among grains. 
There are various ways to supply energy to the system: a way commonly used in experiments is use of a vibrating bottom plate. 
In computer simulations, a stochastic thermal wall is also often used as an ideal model of energy supply from boundaries. 
When the balance between energy input and dissipation is achieved, the system is in a nonequilibrium steady state. 

It is known that the granular fluids under external driving show various fascinating phenomena, such as convection inside the fluids, waves and patterns that appear on the surface, size segregation like the Brazil nut phenomenon, a transition from a condensed state to a fluidized state (See Ref.~\cite{aranson_2006} and references therein). 
It is also important to study how macroscopic quantities in the system, e.g., height of the center of mass (COM) and granular temperature, 
depend on the system parameters, such as the number of particles, the restitution coefficient, the amplitude and frequency of the external vibration. Experimental  studies have revealed some scaling relationships for the macroscopic quantities~\cite{clement93,luding94,warr95,wildman01}.

There have been also a lot of theoretical studies to understand these phenomena and properties. 
At the most microscopic scale, molecular dynamics (MD) simulations that follow the motion of all particles using computer have been commonly used~\cite{clement93,luding94,luding94-2,luding95,mcnamara97,mcnamara98,isobe99,isobe01}. 
One of the most important tools to study dilute granular gas is kinetic theory, which is formulated in terms of the Boltzmann (-Enskog) equation under the assumption of "molecular-chaos". 
It has been used to study velocity distributions and the scaling relationships in granular fluids under external vibration~\cite{warr95,bernu94,kumaran98,kumaran98-2,sunthar-kumaran99,soto04}. At more macroscopic scale, 
hydrodynamic equations which are derived from the Boltzmann equation have been applied to driven granular systems to study local density, flow, and temperature profiles~\cite{haff83,lee95,lee97,brey01,meerson03}. 

These theoretical approaches have successfully described a variety of phenomena mentioned above, within the range of
validity of each approach. It was found, however, some discrepancies in the results of these
approaches, when they were applied to the scaling relationships for the macroscopic quantities.
For example, as summarized in the review of previous works by McNamara and Luding~\cite{mcnamara98}, experimental results by Warr, Huntley, and Jacques~\cite{warr95}, suggest a scaling law 
for the height of the COM $h_{c.m.}$ measured from its height at rest $h_{c.m.0}$ taking the form
\begin{eqnarray}
h_{c.m.}-h_{c.m.0}\sim (A_0\omega)^{\delta}H^{\nu},
\end{eqnarray}
with $\delta=1.3\pm 0.04$ and $\nu=0.27\pm 0.11$, where 
 $A_0\omega$ represents the maximum velocity of the vibrating bottom, and $H$ is the number of layers of
particles at rest. Results of MD simulations of a two-dimensional system 
by Luding, Herrmann, and Blumen~\cite{luding94-2}, however, suggest
$\delta\simeq 1.5$ and $\nu \simeq 1$ for simulations without rotation of particles; 
results by Luding~\cite{luding95} give
$\delta=1.60\pm 0.10$ and $\nu = 0.76\pm 0.11$ for simulations including rotation of particles. 
On the other hand, results of kinetic theoretical approaches~\cite{warr95,kumaran98,kumaran98-2} give $\delta=2$, $\nu=1$. 
Although there are some studies that take into account an effect of the wall~\cite{mcnamara98}, an effect of high density ~\cite{sunthar-kumaran99}, and an effect of nonuniformity of the hydrodynamic fields~\cite{brey01}, 
this discrepancy has not yet been fully explained. 

Recently we have proposed a novel theoretical approach~\cite{woi_2008,wi_2009} which describes the system at more macroscopic level
than the previous approaches mentioned above.
Our theory is on the basis of a Langevin equation of motion of the COM, which is derived by focusing on the 
force exerted to the granular media by the bottom plate. Some macroscopic quantities can be easily  
obtained from the solution of this equation.
In this proceedings, we will first describe how to formulate the Langevin equation for the COM,
and then test the theory by comparing some important predictions derived from the Langevin equation 
with the result of extensive event-driven MD simulations.
A detailed discussion on validity of the fluctuation-dissipation relation which connects the intensity of 
the random force with the friction coefficient will be one of the main points of this article.
 

\section{Theoretical formulation}
In the following we explain our theoretical formulation in 
the case of $D$-dimensional system of grains on a vibrating bottom plate.
The case of grains on a heat source at bottom 
can be discussed only by a slight modification
as shown later.

We consider granular medium that consists of $N$ inelastic particles of mass $m$ and diameter $d$ bouncing on a vibrating bottom plate, subjected to gravity with acceleration $g$. The $z$-direction is chosen to be 
opposite to the direction of gravity. 
The motion of particles is confined in a box of cross section $S$, where the quantity $S$ is an area for $D=3$, a length for $D=2$, and unity for $D=1$.
The bottom plate oscillates sinusoidally with amplitude $A_0$ and angular frequency $\omega$. Hence, 
the height of the bottom plate $z_0(t)$ at time $t$ is
\begin{eqnarray}
z_0(t)=A_0\sin(\omega t).
\end{eqnarray}

In this paper, we focus only on a fluidized state without any structure in the horizontal
directions, such as convection and surface wave. Such a fluidized state 
can be achieved assuming sufficiently small length scales of the cross section of the box containing granular medium, 
so that convections and surface waves are suppressed.      
Moreover, we ignore any boundary effects associated with the side walls for simplicity. Indeed, in event-driven MD simulations to be described 
later, periodic boundary conditions in the horizontal directions have been used.

\subsection{Langevin equation}
We will begin by considering the equation of motion of the COM of the particles:
\begin{eqnarray}
M\frac{d^2 Z}{dt^2}=-Mg+F_b.
\label{eqmotion}
\end{eqnarray}
Here, $M\equiv Nm$ is the total mass of particles and $Z$ is the height of the COM. 
Ignoring the boundary effects concerning the side walls,
the force acting on the COM is the sum of gravity and the force exerted by the bottom plate $F_b$.

There are two important timescales in the system:
the oscillation period of the bottom plate $\tau(\equiv 2\pi/\omega)$ and
the macroscopic relaxation time $\tau_{rel}$ to the stationary state.
If $\tau$ is comparable to $\tau_{rel}$, the energy supplied by one stroke of
the bottom plate is almost dissipated during the period, and the particles
will be in a condensed state. Such a condensed state is beyond the scope of the 
present study. 
In this study, we restrict ourselves to the high-frequency case
$(\tau/\tau_{rel})^2\ll 1$, in which the system is in a fluidized state.

In order to evaluate $F_b$, the force exerted on the particles
by the bottom plate, 
we consider its reaction force, that is, the force exerted on the bottom 
plate by the particles. 
Here we draw an analogy between our system and a system of a Brownian particle exhibiting Brownian motion 
(see, for example, Ref.~\cite{kubo85}).
In our system, granular particles randomly collide with the bottom wall 
and exert a force on it. In Brownian motion, solvent molecules randomly collide with a Brownian particle and exert a force on it.
These two forces are expected to have similar properties. The force on one side of the Brownian particle consists of a force due to the static pressure, the frictional force, and the random force. Thus, on the basis of the analogy, it is plausible to assume that $F_b$ consists of the following four kinds of forces: average force due to the static pressure, frictional force, elastic force, and random force. 

The elastic force has its origin in elastic oscillation excited by the bottom plate. An elastic oscillation mode that shows the slowest relaxation is the mode with the largest-wavelength in the system. In the stationary state, this mode plays a dominant role in deciding macroscopic property of the system. In the largest-wavelength mode, the granular fluid shows macroscopic oscillation alternating between the expansion state with the highest surface of the fluid and the contraction state with the lowest surface, which accompanies oscillation of the height of the COM with the same frequency. It must be noted that the oscillation of the bottom plate does not trigger resonances with the largest-wavelength mode and the other macroscopic modes, because the condition we assumed above, $(\tau/\tau_{rel})^2\ll 1$, suggests that the time scale of the bottom plate oscillation is much shorter than time scales of macroscopic motions. Another important elastic oscillation is a sound wave that is directly excited by the bottom plate. Hence, we take into account two forces as the elastic force: a force due to this largest-wavelength mode and a force due to a sound wave excited by the bottom plate oscillation. 

Finally, we assume the following form of $F_{b}(t)$:
\begin{eqnarray}
F_b(t)=Mg-M\mu V_z (t)-M\Omega^2 \left(Z(t)-\overline{Z}\right)+f_s +R(t).
\label{fb}
\end{eqnarray}
The first term is the average force acting on the bottom plate.
The second term is the frictional force, which is assumed to be linear in the $z$-component of
the velocity $V_z$ of the COM, with a constant coefficient $\mu$. 
The third term is the elastic force resulting from the largest-wavelength mode showing macroscopic expansion and contraction of the granular fluid. 
We assume that it is linear in the height of the center of mass $Z$ measured 
from its long time average $\overline{Z}$. A constant coefficient $\Omega$
specifies the angular frequency of macroscopic oscillation of the granular fluid. 
The fourth term is the other elastic force resulting from excitation of a sound wave by the vibrating bottom plate. 
We will formulate this term below. The last term is the random force. Its property will also be specified later.

The period of the macroscopic slowest oscillation $\tau_{osc}$ is given by 
$\tau_{osc}=2\pi/\Omega$. Similarly, the macroscopic relaxation time $\tau_{rel}$
is given by $\tau_{rel}=1/\mu$. Since the both time scales characterize
macroscopic changes that extend to the whole system,
it is plausible to assume that they are on the same order as the characteristic
time for a sound wave to travel along the vertical direction from the bottom 
to the surface of the granular gas. 
Let us define the thermal velocity $c$,
by $c=\sqrt{D k_B T/m}$, where $T$ is the global temperature 
related to the mean square velocity fluctuation and $k_B$ is the Boltzmann constant.
Then, the characteristic time can be estimated as $c/g$ because the velocity of sound is on the order of $c$ and the height of the surface of the granular gas is,
as the first order approximation, on the order of $c^2/2g$, which is 
the maximum height of a particle launched from the bottom with the 
thermal velocity $c$.
Thus, we assume that $\Omega$ and $\mu$ are on the order of $g/c$:
\begin{eqnarray}
\Omega=\hat{\Omega}\frac{g}{c},\hspace{0.5cm}
\mu=\hat{\mu}\frac{g}{c},
\end{eqnarray}
where $\hat{\Omega}$ and $\hat{\mu}$ are numerical factors on the order of 1
that are determined curve-fitting the results of our simulation.

We estimate the elastic force $f_s(t)$ on the basis of hydrodynamic sound-wave theory~\cite{landau87}.
In a normal fluid, sound waves propagate according to a relationship between the pressure and the velocity of the fluid.
Let us denote a small change in the pressure from its equilibrium value by $p'$, a typical velocity of the fluid particles in the wave by $v$, and the velocity of sound by $c_s$.
If the condition $v \ll c_s$ is satisfied, we have a relationship $p'=\rho c_s v$ for a traveling plane wave, where $\rho$ is the constant equilibrium density of the fluid.
We assume that this relationship is also satisfied in fluidized granular media under the same condition $v\ll c_s$.
The velocity of sound $c_s$ is on the order of the thermal velocity $c$.
The density $\rho$ is on the order of $M/(S c^2/2g)$, where $c^2/2g$ is 
the first order estimation of the surface height of the granular gas 
as described above, and $S$ represents the area of the base
of the system.
In the vicinity of the bottom plate, $v$ may be approximated by the velocity of the bottom plate $v_0(t)=A_0\omega\cos(\omega t)$.
Since $f_s(t)$ corresponds to $p'S$ at the bottom plate, we have
\begin{eqnarray}
f_s(t)=\hat{\sigma}M\frac{g}{c}A_0\omega\cos(\omega t),
\label{fs}
\end{eqnarray}
where $\hat{\sigma}$ is a numerical factor on the order of 1 that is used as a curve-fit parameter when we compare our theoretical predictions with the results of simulations. The condition $v\ll c_s$ is also written as $v_0\ll c$ in the 
vicinity of the bottom plate. Hence, the maximum value of $v_0(t)$, $A_0\omega$,
must be much small compared to $c$: $A_0\omega\ll c$.
Similar estimation of the pressure of sound wave has already been discussed in 
Ref.~\cite{mcnamara97}. 

As property of the random force, we assume stationary Gaussian white noise:
\begin{eqnarray}
\left<R(t)\right>=0,
\hspace{0.5cm}\left<R(t)R(t')\right>=I\delta(t-t').
\label{gaussianwhite}
\end{eqnarray} 
We also assume that the intensity of the random force $I$ is determined by 
the fluctuation-dissipation relation, which is expected to be satisfied when the system is close to equilibrium state: 
\begin{eqnarray}
I=2M\mu k_BT.
\label{fdt2}
\end{eqnarray}
We will discuss later how the fluctuation-dissipation relation
is violated in the stationary state which deviates far
from equilibrium.

Substituting Eq.~(\ref{fb}) into Eq.~(\ref{eqmotion}), we have the following 
linear Langevin equation for the COM:
\begin{eqnarray}
\frac{dV_z }{dt}=-\Omega^2 \left(Z-\overline{Z}\right)-\mu V_z  +
\frac{f_s}{M}+\frac{R}{M}.
\label{lang}
\end{eqnarray}

\subsection{Analytical solution}
It is straightforward to obtain the analytical solution of the Langevin equation
(\ref{lang}) of the form
\begin{eqnarray}
Z(t)-\overline{Z}&=&A_0\zeta\sin(\omega t +\theta)
+\int_{-\infty}^{t}G(t-t')\frac{R(t')}{M}dt'
+F_{ini}(t),
\label{cmheight}
\end{eqnarray}
where
\begin{eqnarray}
\zeta&=&\frac{\hat{\sigma}\,\frac{g}{c}\,\omega}{\sqrt{(\Omega^2-\omega^2)^2+(\mu\omega)^2}},
\label{zeta}
\end{eqnarray}
and
\begin{eqnarray}
\tan\theta &=& -\frac{\omega^2-\Omega^2}{\mu\omega}
\hspace{0.5cm}
\left(-\frac{\pi}{2} \le \theta < 0 \right),
\end{eqnarray}
respectively.
The function $G(t)$ is given by
\begin{eqnarray}
G(t)&=&\frac{e^{-\frac{\mu}{2}t}}{\omega_0}
\sin\left(\omega_0 t\right),
\label{gt}
\end{eqnarray}
where $\omega_0=(\Omega^2-(\mu/2)^2)^{1/2}$.
The last term $F_{ini}(t)$ consists of those that depend on the initial conditions and vanish after a sufficient amount of time. 
Thus, the term is negligible when calculating long-time averages of physical quantities in the stationary state.

\section{Numerical test of the Langevin approach}

We have performed event-driven MD simulations of 
one- and two-dimensional systems.
The one-dimensional system consists of inelastic hard rods on a vibrating
bottom plate. The two-dimensional system consists of inelastic hard disks
on a thermal bottom wall.
These two systems have been investigated in many previous numerical and
theoretical works. 
In our two-dimensional simulations, we have used an efficient 
event-driven MD algorism developed by one of the authors~\cite{isobe99-2}.
Here we will demonstrate how our theory based on 
the Langevin equation can describe macroscopic properties of the granular
fluids.

\subsection{1D system on a vibrating bottom plate}
We first study a one-dimensional granular fluid on a vibrating bottom plate~\cite{clement93,luding94,bernu94}.
We performed simulations systematically by changing the number of particles $N$ ($N=10, 20, 100, 1000$), the restitution coefficient $r$ ($r=0.80\sim 0.9999$), and the maximum acceleration of the bottom plate $\Gamma\equiv A_0\omega^2/g$ ($\Gamma=10\sim 2560$).
The physical quantities are averaged over a sufficiently long period of time.
When we compare our theory with simulation, we evaluate the global temperature $T$ by $k_B T/2= \overline{E}_K$,
where $\overline{E}_K$ is the stationary value of the kinetic energy per particle defined as
\begin{eqnarray}
\overline{E}_K\equiv \lim_{t_M\to\infty}
\frac{1}{t_M}\int_{0}^{t_M} \frac{1}{N}\sum_{i=1}^{N}\frac{m}{2}v_i(t)^2 \,dt,
\end{eqnarray}
where $v_i$ is velocity of the i-th particle. Hence, the thermal velocity $c$ is evaluated as
$c\equiv \sqrt{k_B T/m}=\sqrt{2\overline{E}_K/m}$.

Using the solution of the linear Langevin equation, we can calculate some macroscopic quantities. 
Let us first consider the amplitude $\zeta$ of the oscillation of the COM.
Theoretical prediction is given in Eq.~(\ref{zeta}). 
It must be noted here that we consider a fluidized state with the timescale $\left(\tau/\tau_{rel}\right)^2\ll 1$.
This can be rewritten by substituting $\tau=2\pi/\omega$ and $\tau_{rel}=(c/g)/\hat{\mu}$ as $\hat{\omega}^2\gg 1$, where $\hat{\omega}$ is defined as $\hat{\omega}=\omega c/g$.
In this limit, we expanded $\zeta$ in Eq.~(\ref{zeta}) in terms of $\hat{\omega}^{-2}$:
\begin{eqnarray}
\zeta
&=&\frac{\hat{\sigma}}{\hat{\omega}}
\left(1+O\left(\hat{\omega}^{-2}\right)\right).
\label{zeta2}
\end{eqnarray}
\begin{figure}[h]
\includegraphics[height=7.0cm,clip]{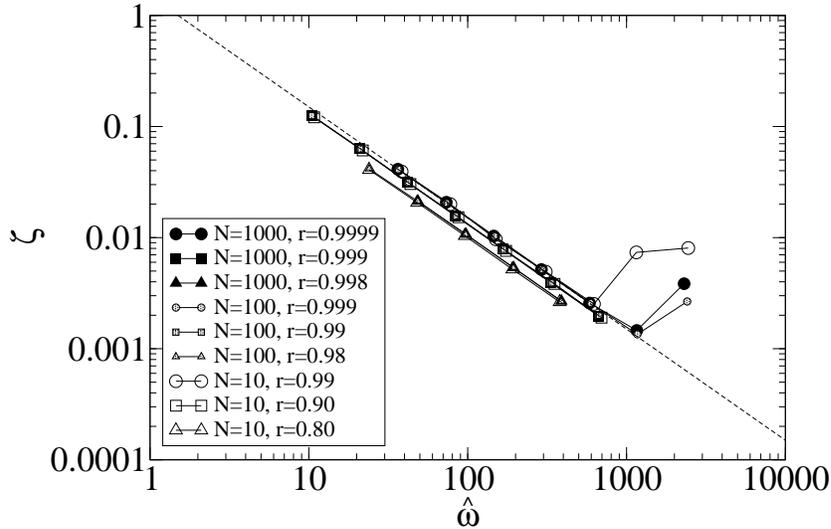}
\caption{\label{fig:zeta}
The amplitude $\zeta$ of the oscillation of the center of mass as a function of the scaled
angular frequency $\hat{\omega}\equiv \omega c/g$ of the bottom plate vibration.
The accelerations $\Gamma$ used in the simulations are $\Gamma=10, 20, 40, 80, 160, 320$, and $640$, except for the simulations that experienced an "inelastic collapse" (an infinite number of collisions in a finite time)~\protect\cite{bernu90,mcnamara92}.
The dashed line corresponds to the theoretical prediction $\hat{\sigma}/\hat{\omega}$
given in Eq.(\ref{zeta2}) with a curve-fit parameter $\hat{\sigma}=1.5$.
}
\end{figure}
Thus, it is predicted that the amplitude is inversely proportional to the scaled angular frequency of the vibration. 
Figure 1 gives the simulation results for $\zeta$ as a function of $\hat{\omega}$.
We obtained good agreement between the theoretical prediction (\ref{zeta2}) and the simulation result
when we chose the curve-fit parameter $\hat{\sigma}=1.5$.

\begin{figure}[h]
\includegraphics[height=7.0cm,clip]{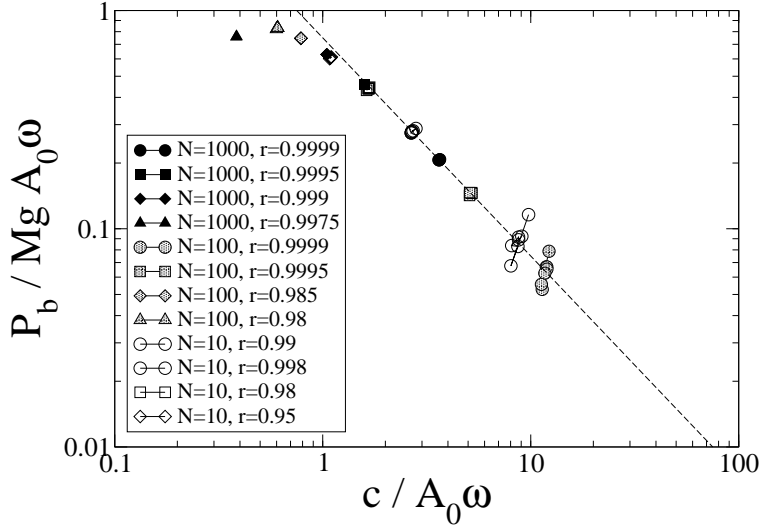}
\caption{\label{fig:scpb} 
The power injection by the bottom plate $P_b$ scaled by $MgA_0\omega$, 
as a function of $c/A_0\omega$. Here $c$ is the thermal velocity evaluated as $c=\sqrt{2\overline{E}_K/m}$. 
The accelerations $\Gamma$ used in the simulations were $\Gamma=10, 20, 40, 80, 160, 320$, and $640$, except the simulations that experienced an inelastic collapse.
The dashed line corresponds to the theoretical prediction $\left(\hat{\sigma}/2\right)\left(A_0\omega/c\right)$ given by Eq.~(\ref{pbsc}) with a curve-fit parameter $\hat{\sigma}=1.5$.
}
\end{figure}

Secondly, we consider the power injection by the bottom plate $P_b$, which has been 
the subject of recent studies~\cite{warr95,mcnamara97,kumaran98,kumaran98-2,soto04}. 
It is defined by using the force exerted by the bottom plate $F_b$ and
its velocity $v_0$:
\begin{eqnarray}
P_b
=\lim_{t_M\to\infty}\frac{1}{t_M}\int_{0}^{t_M}F_b(t)v_0(t)dt.
\label{pbdef}
\end{eqnarray}
Substituting $F_b$ using Eq.~(\ref{eqmotion}) into Eq.~(\ref{pbdef}), and
integrating by parts twice, we have
\begin{eqnarray}
P_b 
= \lim_{t_M\to\infty}\frac{1}{t_M}\int_{0}^{t_M}
\left(M\frac{d^2 Z}{dt^2}+Mg\right) v_0(t)dt
= -M\omega^2
\lim_{t_M\to\infty}\frac{1}{t_M}\int_{0}^{t_M}
Z(t)v_0(t) dt.
\label{pb}
\end{eqnarray}
Then, substituting Eq.~(\ref{cmheight}) into Eq.~(\ref{pb}), we obtain
\begin{eqnarray}
P_b/MgA_0\omega
=\frac{\hat{\sigma}}{2}\frac{A_0\omega}{c}
\frac{\omega^2\left(\omega^2-\Omega^2\right)}
{\left(\omega^2-\Omega^2\right)^2+\left(\mu'\omega\right)^2}
=\frac{\hat{\sigma}}{2}\frac{A_0\omega}{c}
\left(1+O\left(\hat{\omega}^{-2}\right)\right).
\label{pbsc}
\end{eqnarray}
The results obtained by neglecting terms on the order of $\hat{\omega}^{-2}$ coincide with the scaling predicted by kinetic theories~\cite{warr95,kumaran98,kumaran98-2}: $P_b\sim Mg\left(A_0\omega\right)^2/c$. Figure~\ref{fig:scpb} shows that the scaling relationship (\ref{pbsc}) with the same curve-fit parameter as Fig.1, $\hat{\sigma}=1.5$, agrees well with the simulations in the region where $A_0 \omega/c \alt 1$.
This result is consistent with the condition $A_0 \omega/c \ll 1$ required for the formula given by Eq.~(\ref{fs}) to be valid.

Thirdly, let us consider the power spectrum $I_{CM}$ for the height of the COM. 
According to the Wiener-Khinchin theorem, this can be calculated analytically from the Fourier transform of the 
two-time correlation function $\psi_{CM}(t)$ defined by 
\begin{eqnarray}
\psi_{CM}(t)=\lim_{t_M\to\infty}\frac{1}{t_M}\int_{0}^{t_M}
\left< \delta Z(t')\delta Z(t'+t)\right>dt'\,
,
\label{psicm}
\end{eqnarray}
where $\delta Z(t)\equiv Z(t)-\overline{Z}$ and the brackets $\left<\cdots\right>$ indicate an average over the random force $R(t)$.
Substituting Eq.~(\ref{cmheight}) into Eq.~(\ref{psicm}) and performing the Fourier transform, we obtain 
\begin{eqnarray}
I_{CM}(\hat{\omega'})/ \frac{c^5}{Ng^3}
=
\frac{\pi}{2}N\zeta^2
\left(\frac{Ag}{c^2}\right)^2
\left(\delta(\hat{\omega'}-\hat{\omega})
+\delta(\hat{\omega'}+\hat{\omega})\right)
+
\frac{2\hat{\mu}}{(\hat{\Omega}^2-\hat{\omega'}^2)^2+(\hat{\mu}\hat{\omega'})^2}
,
\label{pscmhsc}
\end{eqnarray}
where $\hat{\omega}'$ is the angular frequency $\omega'$ scaled by $g/c$:
$\hat{\omega}'=\omega'c/g$.
Hence, our theory predicts that the power spectrum consists of two terms: 
the first term gives the delta-functional peak at the frequency of the bottom plate oscillation $\hat{\omega}\equiv \omega c/g$. 
The second term represents a continuous spectrum. 
Indeed it has already been shown in Refs.~\cite{clement93,luding94} that the power spectrum for the motion of the COM in a fluidized state
consists of a continuous spectrum and a sharp peak at the frequency of the vibration.
In Fig.3, we present the scaled power spectrum for the case of 
$N=100$, $r$=0.99, and $\Gamma$=160 obtained from our simulation. 
\begin{figure}[h]
\includegraphics[height=7cm,clip]{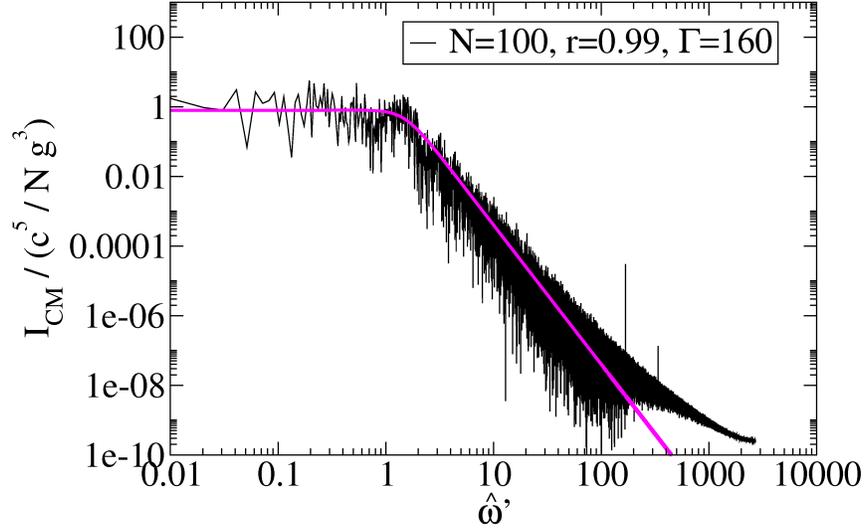}
\caption{\label{fig:pscom-sc} 
The power spectrum $I_{CM}$ for the height of the center of mass scaled by $c^5/Ng^3$
obtained from simulations for $N=100$, 
$r$=0.99, and $\Gamma$=160.
The gray solid line depicts the theoretical prediction given by the second term in Eq.~(\ref{pscmhsc}) with curve-fit parameters $\hat{\mu}=2.0$ and $\hat{\Omega}=1.5$.}
\end{figure}
\begin{figure}
\includegraphics[height=7cm,clip]{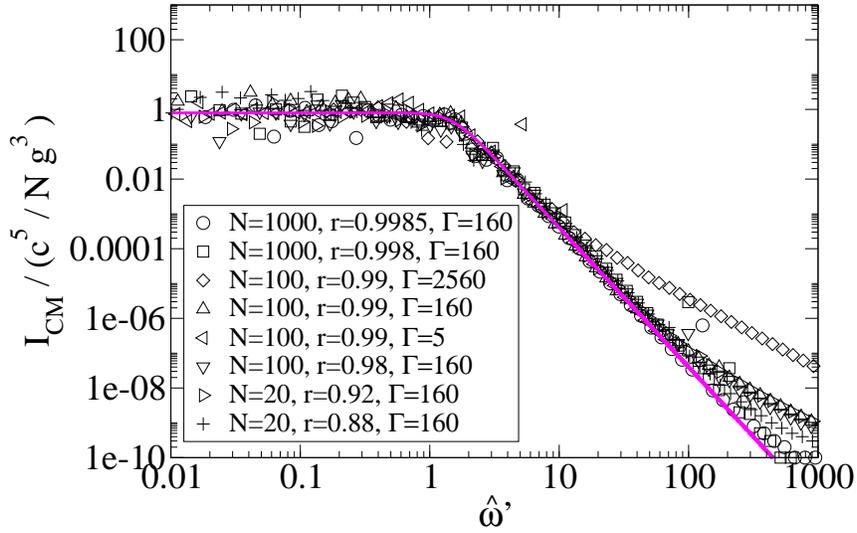}
\caption{\label{fig:pscom-sc2}
The power spectrum $I_{CM}$ for the height of the center of mass scaled by $c^5/Ng^3$
for various system parameters.
The gray solid line is the same as in Fig.3.
}
\label{fig:}
\end{figure} 
The result of simulation indeed shows the sharp peak corresponding to the first term and the continuous part corresponding to the second term.  The part of continuous spectrum agrees well with our theoretical prediction when we choose the curve-fit parameters $\hat{\mu}=2.0$ and $\hat{\Omega}=1.5$.

The continuous part of the scaled power spectrum given by the second term of Eq.~(\ref{pscmhsc}) predicts a universal behavior independent of the system parameters. 
In order to confirm this property, we first divided the logarithm of the angular frequency $\log \omega'$ into bins with a constant interval, and took an average of the power spectrum over the bins; then we scaled it as given by Eq.~(\ref{pscmhsc}).
The results of simulations with various system parameters
are shown in Fig.4.
The simulation data indeed collapse on a single master curve which agrees very well with the curve of the theoretical prediction.

Furthermore, we study how the law of equipartition fails when the inelasticity is increased. 
Let us define the stationary value of the kinetic energy of the COM for the motion along the $z$-axis as 
\begin{eqnarray}
\overline{K}_{CM}\equiv \lim_{t_M\to\infty}\frac{1}{t_M}\int_{0}^{t_M}\left<\frac{1}{2}MV_z(t)^2 \right>dt.
\label{kcmdef}
\end{eqnarray}
Using Eq.~(\ref{cmheight}), $\overline{K}_{CM}$ can be calculated and yields
\begin{eqnarray}
\frac{\overline{K}_{CM}}{\overline{E}_{K}}=1+\frac{N}{8}\left(\frac{mgA_0\hat{\sigma}}{\overline{E}_{K}}\right)^2.
\label{kcm}
\end{eqnarray}

Figure 5 shows comparison of simulation results with the theoretical prediction (\ref{kcm}). 
Here, the vertical axis shows the ratio of the kinetic energy of the COM, 
$\overline{K}_{CM}$, to the kinetic energy per particle $\overline{E}_K$. When the law of equipartition is satisfied, this ratio is equal to 1. The simulation data show deviation from the equipartition at $\overline{E}_K\ll 10$. This deviation can be well explained by our theory as far as it is not too large, using 
the same curve-fit parameter $\hat{\sigma}=1.5$ as used in Fig.1 and Fig.2.

\begin{figure}[h]
\includegraphics[height=7.0cm,clip]{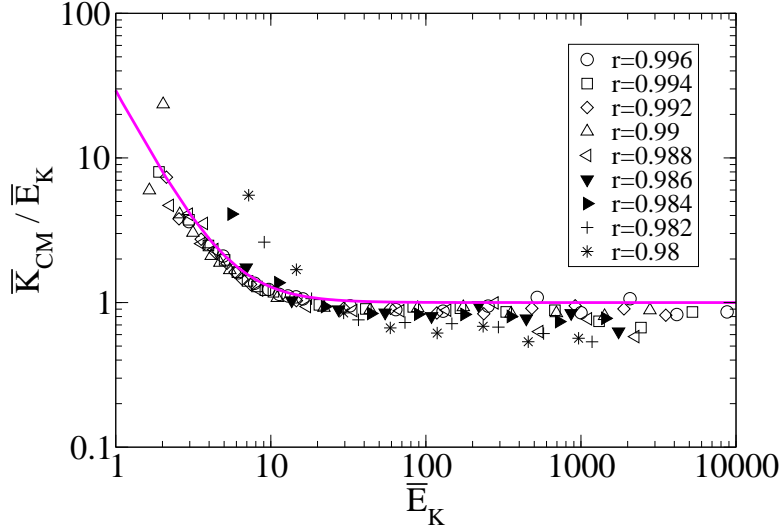}
\caption{\label{kratio} 
The ratio of kinetic energy of the center of mass $\overline{K}_{CM}$ and kinetic
energy per particle $\overline{E}_{CM}$ for $N=100$ and different $\Gamma$ values
(from 10 to 2560),
except the simulations that experienced an inelastic collapse.
The gray solid line gives the theoretical prediction Eq.~(\ref{kcm}) with a curve-fit parameter $\hat{\sigma}=1.5$.
}
\end{figure}

\subsection{2D system on a thermal wall}
We then study a two-dimensional granular fluid on a thermal bottom wall~\cite{isobe99,isobe01,meerson03}.
The system considered here consists of $N$ inelastic hard disks with mass $m$ and diameter $d$, 
in a two-dimensional space of a width $L$. We set periodic boundary condition in the horizontal direction and infinite height in the vertical direction.
The stochastic thermal wall with the temperature $T_0$ is set at the bottom.
When a disk collides with the wall, 
it comes off with the value of the vertical velocity component, 
$v_n$, sampled from the probability density
\begin{eqnarray}
p(v_n)=\frac{mv_n}{k_B T_0}\exp\left(-\frac{mv_n^2}{2k_B T_0}\right).
\end{eqnarray}
In order to prevent unphysical flows in the horizontal direction 
which may be caused by the periodic boundary conditions, 
we impose that the velocity component parallel to the wall remains 
unchanged by the collision with the wall. 
The dynamics is evolved by binary inelastic collisions between disks 
with a constant normal restitution coefficient $r$.
The system can be completely characterized by the number of disks $N$, the
dimensionless width $\hat{L}\equiv L/d$, the dimensionless driving intensity
$\Lambda\equiv k_B T_0/mgd$, and the restitution coefficient $r$~\cite{isobe99,isobe01}. 
The value of $\hat{L}$ is chosen sufficiently small to prevent the density instability, so that the system remains homogeneous in the horizontal direction.

Our theory can be easily applied to the case of the thermal bottom wall
by simply modifying $f_s(t)=0$ in Eq.~(\ref{lang}).
The other difference from the one-dimensional system is that the thermal velocity $c$ is defined by
$c\equiv \sqrt{2k_B T/m}$.
When we compare our theory with simulation, we evaluate $T$ by $k_B T=\overline{E}_K$ and 
$c$ by $c=\sqrt{2\overline{E}_K/m}$. 

Then, the theoretical curve of the power spectrum $I_{\rm CM}$ for the height of the COM is given by the second term in Eq.~(\ref{pscmhsc}) divided by 
the system dimensionality 2, which comes from the change 
in the relation between $k_B T$ and $\overline{E}_K$:
\begin{eqnarray}
I_{CM}(\hat{\omega'})/ \frac{c^5}{Ng^3}
&=&
\frac{\hat{\mu}}{(\hat{\Omega}^2-\hat{\omega'}^2)^2+(\hat{\mu}\hat{\omega'})^2}
.
\label{pscmhsc-2d}
\end{eqnarray}

In the Figs. 6 and 7, the power spectra $I_{\rm CM}$ of the COM with fixed parameters $(\hat{L}, \Lambda)=(10,10^3)$ for various ($N, r$) are shown. 
Figure 6 shows the results of the simulations in the case of nearly elastic collisions ($1-r\ll 1$). Here we find that the scaled power spectra concentrate into a master curve.
The theoretical curve with the curve-fit numerical factors $\hat{\mu}=2.0$ and 
$\hat{\Omega}=1.5$ shows good agreement with numerical simulations.
On the other hand, Fig.7 shows the result of the simulations in the large 
inelasticity case; it demonstrates a systematic deviation 
from the curve predicted by our theory. 
We speculate that this deviation is caused by violation of 
the fluctuation-dissipation relation, which will be discussed
in the following sections.
\begin{figure}[h]
\includegraphics[height=7cm,clip]{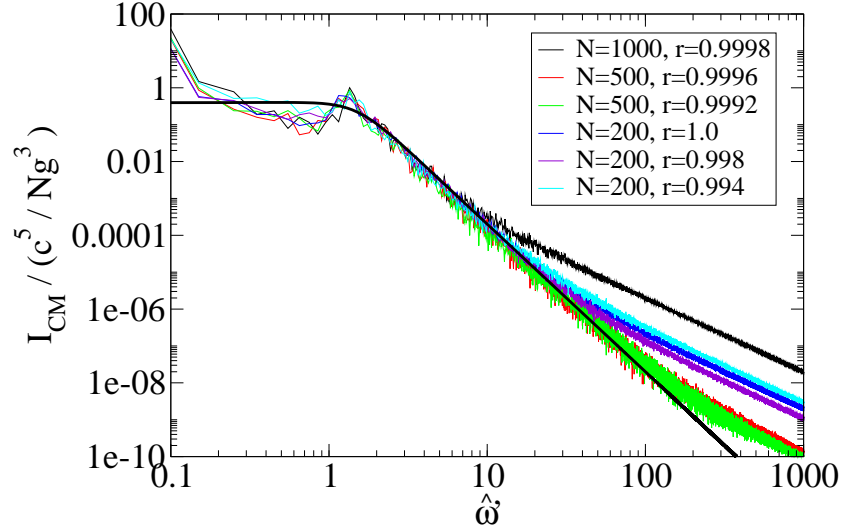}
\caption{\label{2d-pscom-sc} 
The power spectrum $I_{CM}$ for the height of the center of mass scaled by $c^5/Ng^3$ obtained from simulations for various ($N, r$). The driving intensity is $\Lambda=1000$.
Here the restitution coefficients $r$ used in the simulations are close to $1$ (nearly elastic case).
The thick solid line depicts the theoretical prediction given by Eq.~(\ref{pscmhsc-2d}) with curve-fit parameters $\hat{\mu}=2.0$ and $\hat{\Omega}=1.5$.}
\end{figure}
\begin{figure}
\includegraphics[height=7cm,clip]{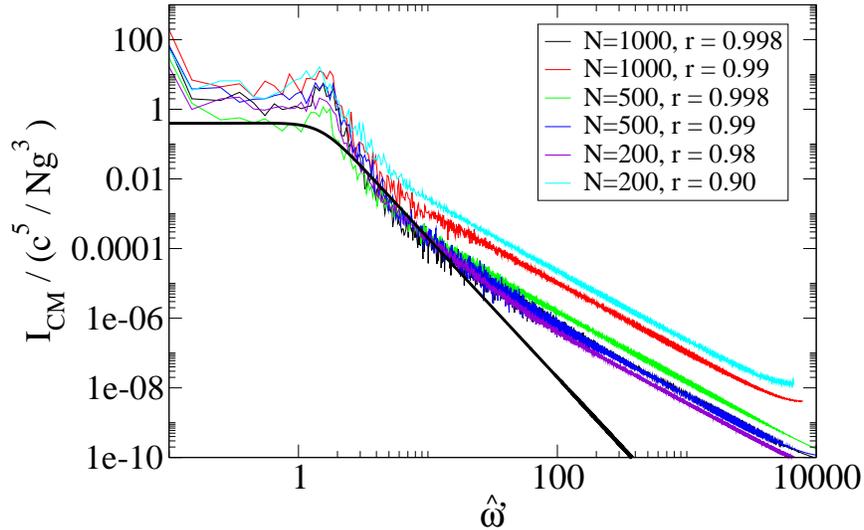}
\caption{The same result as in Fig.6 but the restitution coefficients $r$ are smaller
than those used in Fig.6 (large inelasticity case).
The thick solid line depicts the theoretical prediction given by Eq.~(\ref{pscmhsc-2d}) with curve-fit parameters $\hat{\mu}=2.0$ and $\hat{\Omega}=1.5$.
}
\label{scpscomfel}
\end{figure} 

\subsection{Failure of the law of equipartition in 2D system}
We first investigate how the law of equipartition fails when inelasticity is increased. Figure 8 shows the ratio of the kinetic energy of the COM $\overline{K}_{CM}$ to $k_B T/2(=\overline{E}_K/2)$. 
Since $\overline{K}_{CM}$ is defined by Eq.~(\ref{kcmdef}) using only the $z$-component of the COM velocity $V_z$, 
this ratio is to be equal to 1 when the law of equipartition is satisfied.
We found 
a systematic deviation from the law of equipartition at low temperature;
the deviation seems to be inversely proportional to $T$ and independent of $N$. 
Our theory can not give any explanation on this behavior of the ratio of the energies.  
\begin{figure}[h]
\includegraphics[height=7.0cm,clip]{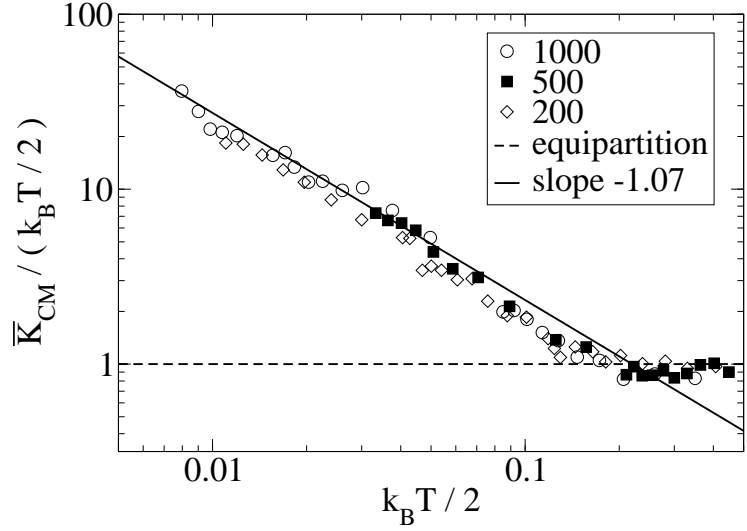}
\caption{\label{kratio} 
The ratio of kinetic energy of the center of mass $\overline{K}_{CM}$ and 
the half of kinetic energy per particle $k_B T/2$ plotted against $k_B T/2$.
Here $k_B T$ is evaluated by $k_B T=\overline{E}_K$.
The restitution coefficients $r$ are from 0.964 to 0.9995 for $N=1000$,
from 0.991 to 0.9999 for $N=500$, and from 0.90 to 0.999 for $N=200$.  
The dashed line gives the law of equipartition of energy.
The solid line gives a slope of $-1.07$. 
}
\end{figure}

It is known from previous studies that the system shows density inversion in a certain condition~\cite{isobe99,isobe01,brey01,meerson03}. 
Here we discuss on a connection between the density inversion phenomena and the failure of the energy equipartition. 
In the previous study based on hydrodynamic equations, 
Bromberg et al.~\cite{meerson03} have shown that if a parameter $\lambda$ defined by 
\begin{eqnarray}
\lambda\equiv \frac{\pi^{1/2}}{2}N_h\left(1-r^2\right)^{1/2},
\label{lambda}
\end{eqnarray}
where $N_h$ is the number of layers of disks at rest,
becomes larger than the critical value $\lambda_c\simeq 1.06569...$,
then density inversion develops in the system.
Here we show in Fig. 9 the ratio of the energies as a function of $\lambda$. 
We observe indeed that the deviation from the equipartition starts at a certain value of $\lambda$ near $\lambda_c$. We also confirmed that at the data points where the law of equipartition fails, the density inversion indeed develops as shown in Fig.10.
\begin{figure}[h]
\includegraphics[height=7.0cm,clip]{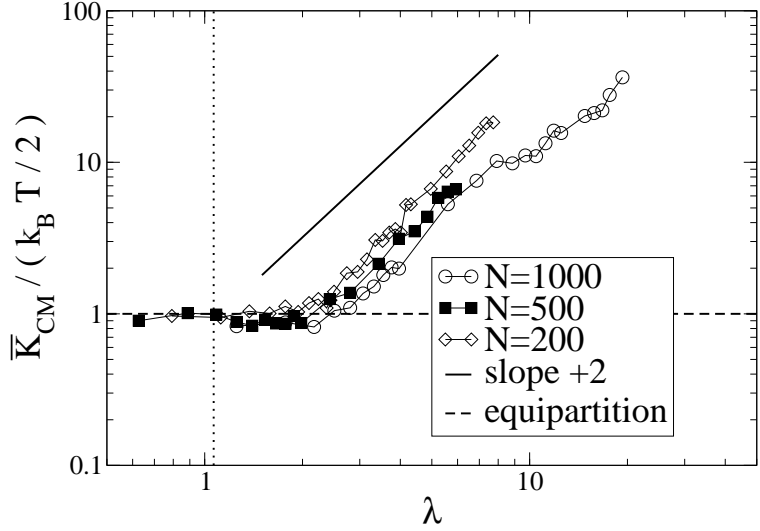}
\caption{\label{kratio} 
The ratio of kinetic energy of the center of mass $\overline{K}_{CM}$ and 
the half of kinetic energy per particle $k_B T/2$ plotted against $\lambda$
defined by Eq.~(\ref{lambda}).
The dashed line gives the law of equipartition of energy.
The dotted line indicates $\lambda_c$, the critical value of $\lambda$
for the density inversion.
The solid line gives a slope of $+2$. 
}
\end{figure}

\begin{figure}[h]
\begin{tabular}{ccc}
\begin{minipage}{0.3\hsize}
\includegraphics[height=9.0cm,clip]{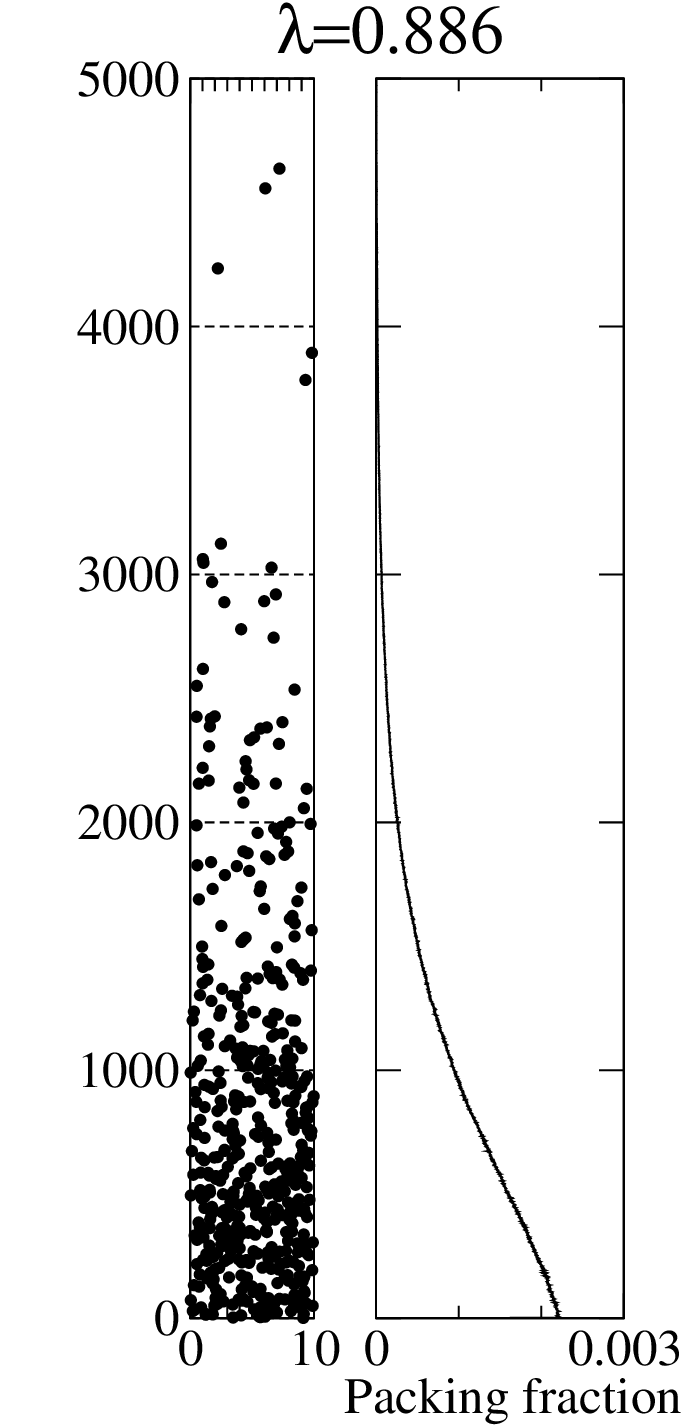}
\end{minipage}
\hspace{0.3cm}
\begin{minipage}{0.3\hsize}
\includegraphics[height=9.0cm,clip]{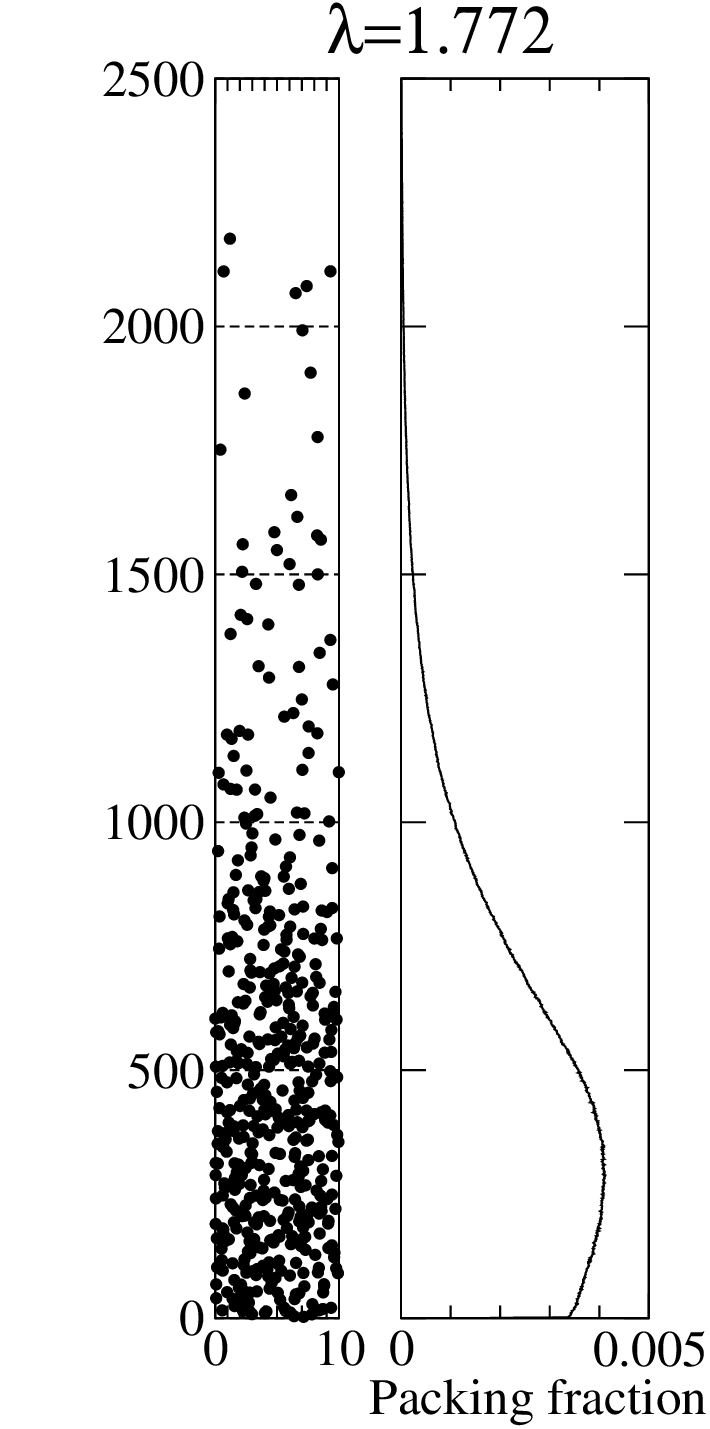}
\caption{\label{snapshot} 
}
\end{minipage}
\hspace{0.3cm}
\begin{minipage}{0.3\hsize}
\includegraphics[height=9.0cm,clip]{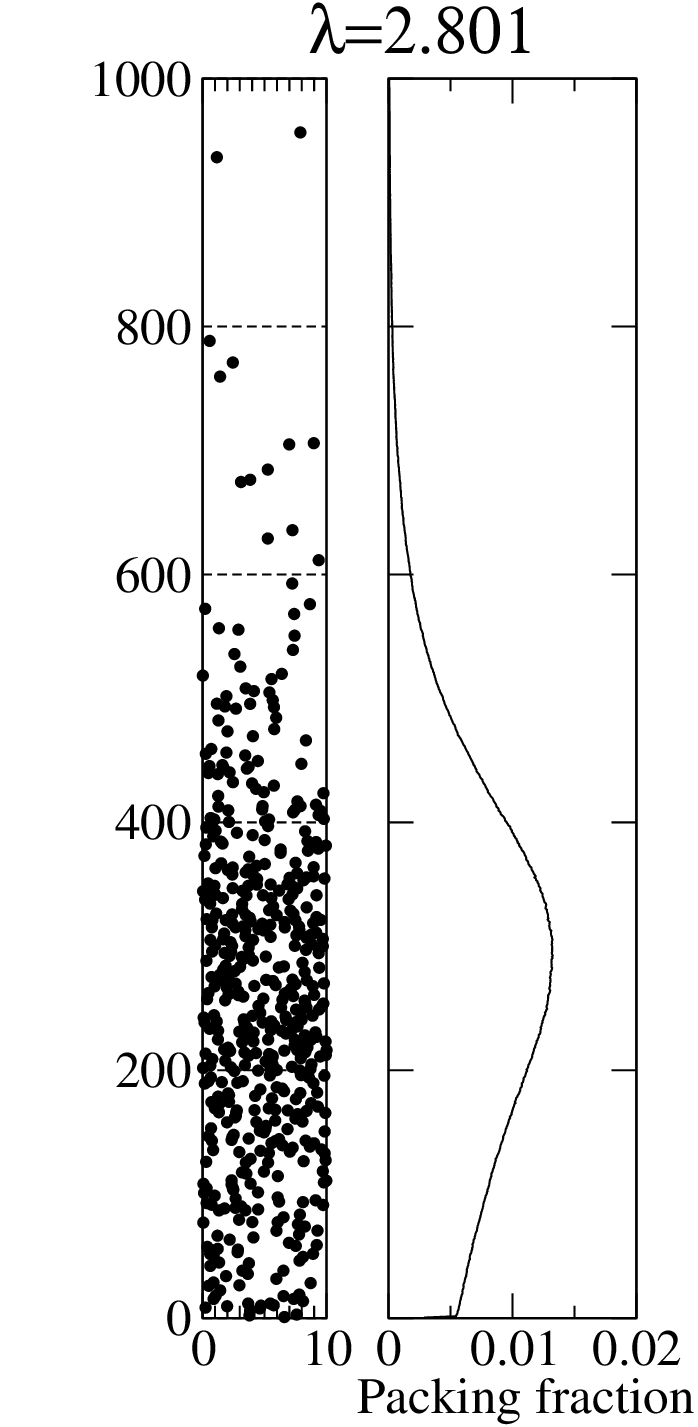}
\caption{\label{snapshot} 
Snapshots and packing fraction profiles of simulations for $\Lambda=1000$, $N=500$, and 
different $\lambda$ values.
}
\end{minipage}
\end{tabular}
\end{figure}

\subsection{Failure of fluctuation-dissipation relation in 2D system} 
The fluctuation-dissipation relation (of the second kind) (\ref{fdt2}) 
with respect to our Langevin equation (\ref{lang}) with $f_s=0$ can be derived using the
same procedure in the study of Brownian motion based on the Langevin
equation~\cite{kubo85}. To begin with, let us define the intensity of the 
random force $I$ by Eq.~(\ref{gaussianwhite}). Then, we calculate $\overline{K}_{CM}$ using the solution 
(\ref{cmheight}) of
the Langevin equation (\ref{lang}) with $f_s(t)=0$, and represent it in terms of $I$:
\begin{eqnarray}
\overline{K}_{CM}=\frac{I}{4M\mu}.
\label{kcm_1}
\end{eqnarray}
Finally, if the equipartition of energy, that is , $\overline{K}_{CM}=k_B T/2$,
is satisfied, then we obtain the fluctuation-dissipation relation 
as given in Eq.~(\ref{fdt2}).
Therefore, in the case of large inelasticity where the law of equipartition of energy fails, the fluctuation-dissipation relation also fails to be satisfied. 

Having observed that the fluctuation-dissipation relation is violated
in the large inelasticity case, one can then go on to consider 
whether the linear Langevin equation itself is valid in the case of 
large inelasticity.
In order to clarify this point, we use Eq.~(\ref{kcm_1}) and define
 $I$ in terms of $\overline{K}_{CM}$ without use of the fluctuation-dissipation relation: 
$I\equiv 4M\mu \overline{K}_{CM}$.
Then, the theoretical expression of $I_{CM}$ becomes 
\begin{eqnarray}
I_{CM}(\hat{\omega'})/ \left[4\left(\frac{c}{g}\right)^3\frac{\overline{K}_{CM}}{M}\right]
&=&
\frac{\hat{\mu}}{(\hat{\Omega}^2-\hat{\omega'}^2)^2+(\hat{\mu}\hat{\omega'})^2}
.
\label{pscmhsc-2d_2}
\end{eqnarray}
When we compare Eq.~({\ref{pscmhsc-2d_2}) with the results of simulation,
we determine $\overline{K}_{CM}$ from the simulation data.
Figure 11 shows the rescaling of the power spectra successfully collapses
the simulation data onto a single curve near the peak.
The curve of the theoretical expression (\ref{pscmhsc-2d_2}) with
$\hat{\mu}=0.6$ and $\hat{\Omega}=1.5$ seems to show better fitting 
than the case of $\hat{\mu}=2.0$ and $\hat{\Omega}=1.5$.
This change in the coefficient $\hat{\mu}$ in the case of large inelasticity
may be attributed to the internal structure caused by the density inversion.
We have not yet understand how the values of the phenomenological constants
in our theory are related to the density profile of the system.

\begin{figure}[h]
\includegraphics[height=7.0cm,clip]{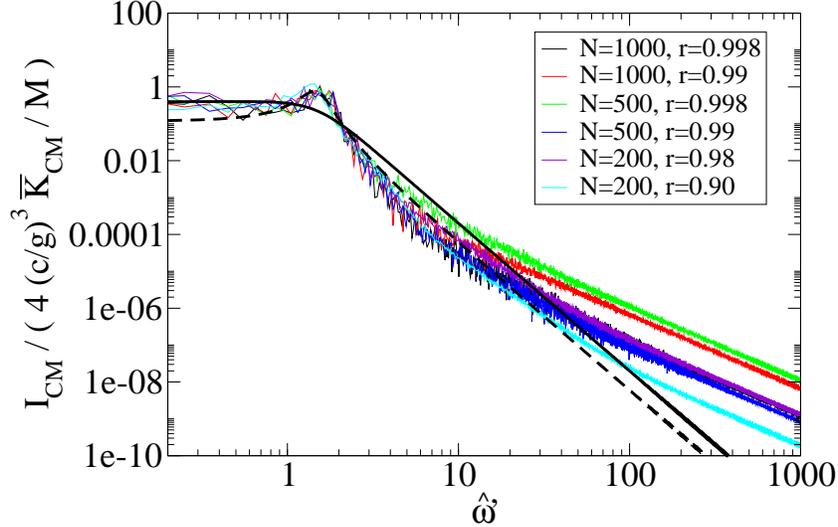}
\caption{\label{psrescaled}
The power spectrum $I_{CM}$ for the height of the center of mass rescaled by 
$4(c/g)^3\overline{K}_{CM}/M$ for various ($N, r$). 
The parameters and lines of the simulations are the same as in Fig.7.
The thick solid line gives the theoretical prediction given by Eq.~(\ref{pscmhsc-2d}) with curve-fit parameters $\hat{\mu}=2.0$ and $\hat{\Omega}=1.5$.
The thick dashed line gives the theoretical prediction with 
$\hat{\mu}=0.6$ and $\hat{\Omega}=1.5$.}
\end{figure}

\section{Concluding remarks}
A Langevin equation for the motion of the center of mass was formulated to describe
macroscopic properties of the fluidized state of granular matter under gravity.
The analytical expressions of some macroscopic quantities were derived and compared
with the results of extensive simulations. 
There are three numerical factors, $\hat{\sigma}$, $\hat{\mu}$ and $\hat{\Omega}$, that can not be specified by our phenomenological consideration; they were used as curve-fit parameters when we compared our theoretical predictions with the results of simulations.
In the present study, we performed event-driven molecular dynamics simulations for 
the following two systems: a one-dimensional system of inelastic rods on a vibrating 
bottom plate and a two-dimensional system of inelastic disks on a thermal bottom wall.

In the one-dimensional system, we found good agreement between our theory and simulation.
The results of simulations suggest the following values of the numerical factors: $\hat{\sigma}=1.5$, $\hat{\mu}=2.0$ and $\hat{\Omega}=1.5$.
Deviation from the law of equipartition is observed when the kinetic energy per particle
is sufficiently small. This deviation is explained well by our theory as far as it is
not too large to violate the fluctuation-dissipation relation.

In the two-dimensional system, the results of simulations are in good agreement with 
the theoretical predictions if the collisions between disks are nearly elastic.
The results of simulations suggest the following values of the numerical factors: $\hat{\mu}=2.0$ and $\hat{\Omega}=1.5$.
As the inelasticity of the interparticle collisions increases, however, 
the power spectrum for the center of mass obtained by the simulations gradually deviates from the prediction of theoretical curve. We speculated that this deviation is caused by violation of the fluctuation-dissipation relation. From the systematic event-driven simulation for the wide range of parameters, we found failure of the law of equipartition of energy with respect to the kinetic energy of the center of mass, which necessarily causes violation of the fluctuation-dissipation relation.
Contrary to the one-dimensional system, the deviation from 
the law of equipartition can not be explained within the framework of our theory.   
Furthermore, we obtained evidence that this failure of the law of equipartition of energy is caused by 
the inversion of the density profile~\cite{isobe99,isobe01,brey01,meerson03}. 
Finally, we showed that the theoretical predictions obtained without assuming the fluctuation-dissipation relation agree with the results of simulations by choosing the curve-fit parameters $\hat{\mu}=0.6$, $\hat{\Omega}=1.5$. This result suggests that the Langevin equation itself might remain valid even when the fluctuation-dissipation relation is violated, and it might be still useful to explain a universal behavior of the power spectrum in the large inelasticity case.

\section{Acknowledgment}

This study was supported by Grant-in-Aid for Scientific Research from the Ministry of Education, Culture, Sports, Science and Technology No. 19740236.
Part of the computations for this study was performed using the facilities of the Supercomputer Center, Institute for Solid State Physics, the University of Tokyo.
This study was financially supported by the Hosokawa 
Powder Technology Foundation.


\end{document}